\newcommand{\myTitle}{Correct Composition of Dephased Behavioural Models}

\newcommand{\Nat}{{\mathbb N}}

\newcommand{\hol}{higher-order logic}

\newcommand{\I}{Isabelle}

\newcommand{\map}[1]{{#1}^{\to}}
\newcommand{\query}[1]{}
\renewcommand{\query}[1] {\marginnote{\raggedright\footnotesize\itshape\hrule\smallskip{#1}\smallskip\hrule}}

\newcommand{\z}{Z3}

\newcommand{\dc}{\downarrow\hspace*{-0.1cm}}

\documentclass
{llncs}

\usepackage[]{cancel}
\usepackage{hyperref}
\usepackage[]{listings}
\usepackage{marginnote}
\usepackage{graphicx,amssymb}
\usepackage{amsmath}
\usepackage{mathtools}
\usepackage{microtype}
\usepackage{listings}
\usepackage{color}

\lstdefinelanguage{Isabelle}[basicstyle=\ttfamily,keywordstyle=\color{blue}\bfseries]{ML}%
{morekeywords={%
assume,%
assumes,%
by,%
def,%
fix,%
fixes,%
from,%
have,%
lemma,%
obtain,%
proof,%
qed,%
show,%
shows,%
theorem,%
try,%
unfolding,%
using,%
where,%
},
stringstyle=\togglelst@instring\lst@eat,
basicstyle=\itshape,%
keywordstyle=\upshape\bfseries,%
columns=fullflexible,%
showstringspaces=false,%
mathescape,%
}[keywords,comments]

\newtheorem{defn}{Definition}

\DeclareMathOperator{\sel}{isSelected}
\DeclareMathOperator{\clock}{clock}
\DeclareMathOperator{\score}{priority}

\DeclareMathOperator{\Score}{Score}

\DeclareMathOperator{\ran}{Range}

\hypersetup{%
pdftitle={\myTitle}%
, pdfauthor={Marco B. Caminati}%
, pdfstartview=FitBH
, pdffitwindow=true
, pdfkeywords={first-order logic, formal methods, SMT, constraint programming, formal verification, theorem proving, Isabelle, SAT solvers, digital healthcare, event structures, traces of execution, optimisation}
, pdflang=en-GB
, pdfstartview= 100 100 10000
, pdfsubject={formal methods}
, pdfwindowui=true
, colorlinks=true
, linkcolor=blue
, urlcolor=blue
, citecolor=red
, breaklinks = true
, menucolor=brown
}

\title{\myTitle{}\thanks{This research is supported by EPSRC grant EP/M014290/1.
}}

\author{\href{http://orcid.org/0000-0002-5918-9114}{Juliana~Bowles} \and{}\href{http://orcid.org/0000-0002-4529-5442}{Marco~B.~Caminati}}

\institute{School of Computer Science, University of St Andrews\\
KY16 9SX St Andrews, United Kingdom\\
\email{\{jkfb$\mid$mbc8\}@st-andrews.ac.uk}
}

\begin{document}
\maketitle
\begin{abstract}
Scenarios of execution are commonly used to specify partial behaviour and interactions between different objects and components in a system. To avoid overall inconsistency in specifications, various automated methods have emerged in the literature to compose (behavioural) models. In recent work, we have shown how the theorem prover Isabelle can be combined with the constraint solver Z3 to efficiently detect inconsistencies in two or more behavioural models and, in their absence, generate the composition. 
Here, we extend our approach further and show how to generate the correct composition (as a set of valid traces) of dephased models.
This work has been inspired by a problem from a medical domain where different care pathways (for chronic conditions) may be applied to the same patient with different starting points.
\end{abstract}

\section{Introduction}
To cope with the complexity of modern systems, design approaches combine a variety of languages and notation to capture different aspects of a system, and separate structural from behavioural models.
In itself behavioural modelling is also difficult, and rather than attempt to model the complete behaviour of a (sub)system \cite{UBC09}, it is easier to focus on several possible scenarios of execution separately. 
Scenarios give a partial understanding of a component and include interactions with  other system components.
In industry, individual scenarios are often captured using UML's sequence diagrams \cite{UML}.
Given a set of scenarios, we then need to check whether these  are correct and consistent, and to do so we first need to obtain the combined overall behaviour. 
The same ideas apply if  we model (partial) business processes within an organisation, for instance using BPMN \cite{bpmn}. 
In either case, we need a means to compose models (scenarios or processes), and when this cannot be done, detect and resolve inconsistencies.

Composing systems manually can only be done for small systems.
As a result,  in recent years, various methods for automated model composition have been introduced  \cite{araujo2004modeling,bowles2007formalAndBordbar,france2004uml,klein2006semantic,liang2008general,rubin2008,whittle2006composing,widl2012guided,zhang2009,BABC2015,BBA2015,BBA2016}.
Most of these methods involve introducing algorithms to produce a composite model from simpler  models originating from partial specifications and assume a formal underlying semantics \cite {klein2006semantic}. 
In our recent work \cite{BABC2015,BBA2015,BBA2016}, we have used constraint solvers for automatically constructing the composed model. 
This involves generating all constraints associated to the models, and using an automated solver to find a solution (the composed model) for the conjunction of all constraints.
We used Alloy  \cite{softwareAbstractions} in \cite{BABC2015,BBA2015} and Z3 \cite{z3} in \cite{BBA2016}. 
We conducted several experiments in  \cite{BBA2016}, showing that Z3 performs much better than Alloy for large systems. 
Using Alloy for model composition, mostly in the context of structural models, is an active area of research \cite{rubin2008,zhang2009}, but the use of Z3 is a novelty of \cite{BBA2016}. 
Even though we used  Z3 in \cite{BBA2016}, we did not explore Z3's arithmetic capabilities, nor did we 
deal with incompatible constraints. 
We have addressed both points more recently in \cite{BC2016}. 

As in our earlier work, our approach in \cite{BC2016} used  event structures \cite{WinNie95} as an underlying semantics for sequence diagrams in accordance to \cite{Kue-TCS-06,Bow06}, and explored how the theorem prover Isabelle \cite{Isabelle-HOL} and constraint solver Z3 \cite{z3} could be combined to detect and solve partial specifications and inconsistencies over event structures. 
In this paper, we go one step further in improving the process of automatically generating correct composition models (through a set of valid traces) for behavioural models that may contain 
inconsistencies. 
We introduce a notion of \emph{dephased} models prior to composition, to make it possible to combine models which do not  start execution simultaneously, and where the \emph{pace} of execution or a notion of \emph{priority} in each model may be different as well. 
The effect is a reduction of detected inconsistencies (if any), and the automated generation of what are valid context-specific traces of execution. 
This work has been inspired by a problem from a medical domain where different care pathways (for chronic conditions) may be applied to the same patient with different starting points (diagnosis).
As an additional contribution, we present in Section~\ref{iz3-sec} an original, general method to provide formal correctness proof for SMT code.


This paper is structured as follows. 
The motivation and contribution of the work presented here are discussed
in Section~\ref{back-sec}, while in Section~\ref{LES-sec} we recall our formal model (labelled event structures).
Section~\ref{iz3-sec} describes how \I{} and Z3 are combined to compute valid traces of execution in specific settings. 
We describe the role that Isabelle plays in our work in Section~\ref{ver-sec}. 
We conclude the paper with a description of related work in Section~\ref{rel-sec}, and a discussion of 
future work in Section~\ref{con-sec}.

\section{Context and Contribution}\label{back-sec}
\label{RefSectTools}
Continuing the work started in \cite{BC2016}, we exploit the interface between \I{} and Z3 to obtain a versatile tool for the specification, analysis and computation of the behaviour of complex distributed concurrent systems. By specifying our partial behavioural models in \I{} we can check automatically their correctness, obtain their composition (if it exists) and fill any gaps, while being able to prove at any point that the models are valid \cite{BC2016}. 
If our model contains inconsistent behaviours, we are able to locate the conflicting events. 
However, we argue in this paper, that we may be overlooking valid behaviour in some cases, and we explore an approach to fine-tune the detection of inconsistent behaviour further. 
In order to do so we allow models to be \emph{dephased}, that is, different scenarios (or similarly for processes) may start execution at different times and continue execution at a different pace. We also consider a notion of priority of (locations in) a model.
We develop a technique to find valid traces by defining exactly how the different scenarios come together  (i.e., how they are dephased) and which traces are closer to satisfying assumed model priorities.

The problem we are addressing has been inspired by a problem from a medical domain where different care pathways 
(essentially processes or behavioural models)
for chronic conditions  are being applied to the same patient, such that:
\begin{itemize}
\item different pathway steps are executed at a different \emph{pace}. For instance, for one condition we may need observations to be carried out every month, whereas for others every three months is sufficient.
\item one of the conditions may be prevalent, in other words, has higher \emph{priority}.
\item some of the possible medications prescribed at a given \emph{step} in the pathway may have higher \emph{priority} due to better treatment effectiveness. For instance, the use of metformin in the treatment of type2 diabetes.
\item the diagnosis of different conditions for a patient are likely to have occurred at different times. For instance, the diagnosis of chronic kidney disease often follows (and may be a consequence of) an earlier diagnosis of type 2 diabetes.  
This leads to the corresponding care pathways starting execution at different times, in other words, their execution is \emph{dephased}. 
\end{itemize}

In particular, having an automated technique that allows us to find valid combined traces taking into account priorities is useful as it gives us a flexible mechanism to identify \emph{different solutions} in similar but different cases. For instance, patients with the same conditions overall but with different orders of diagnosis or prevalent condition. To keep the presentation of this paper more focused, we omit the medical details and instead show how the approach works for an abstract example. Consider the following example using UML sequence diagrams \cite{UML}.

\begin{figure}[h]
\includegraphics[scale=.7]{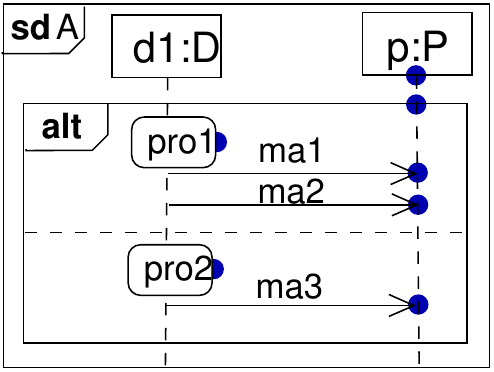}
\hfill
\includegraphics[scale=.7]{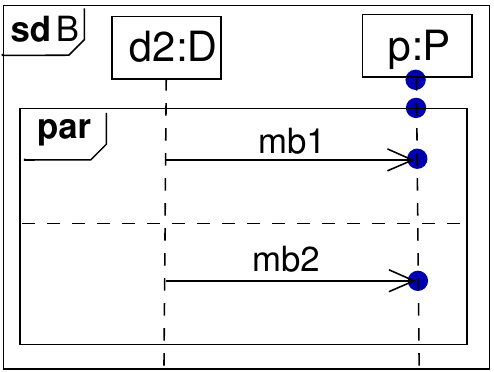}
\hfill
\includegraphics[scale=.7]{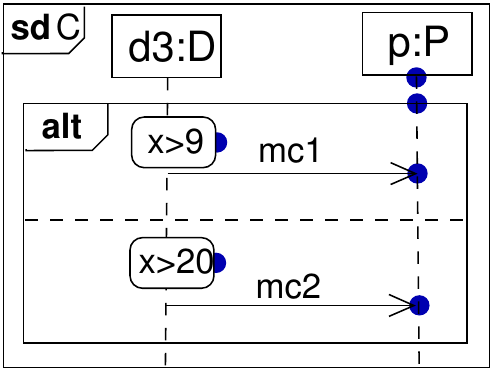}
\caption{Three scenarios involving the same object instances.}
\label{sdia-ex1-fig}
\end{figure}
Fig.~\ref{sdia-ex1-fig} shows three scenarios involving the same instance {\tt p} and different instances of the same class {\tt D}, that is, {\tt d1}, {\tt d2}, {\tt d3}.
The scenarios use interaction fragments for alternative behaviour (indicated by an {\tt alt} on the top-left corner) and parallel behaviour (indicated by a {\tt par}
on the top-left corner). Other fragment operators exist but are not used in this paper (cf. \cite{UML} for details). 
Interaction fragments contain one or more operands, which in the case of an alternative may be preceded by a constraint or guard. 
The 
alternative fragment in {\tt sdA} uses two constraints for the operands, namely {\tt pro1} and {\tt pro2}, and we note that they are not necessarily mutually exclusive.
We may want to associate a priority to {\tt pro1}, to indicate for instance that if it holds we will want the corresponding operand to execute (instead of the second operand and regardless of whether {\tt pro2} holds or not). UML does not have direct notation to indicate this, but we can assume the existence of a priority tag (not shown) and we will add a priority notion to our formal model.
For the messages shown (for instance, {\tt ma1}, {\tt mb1}, {\tt mc1}, and so on), we assume that when they are received, they imply an occurrence for instance {\tt p}. 
The marked points along the lifelines and next to the conditions are what we call \emph{locations}, borrowing terminology from Live Sequence Charts (LSCs) \cite{HarMar03-book}. They do not serve a purpose at the design level but make it easier to understand the formal semantics (cf. \cite{Kue-TCS-06} for details).

Assume that we know that the occurrence of {\tt ma1} conflicts with {\tt mc1}, and
{\tt ma2} conflicts with {\tt mb2}. This is not encoded directly in the scenarios above, but is domain knowledge contained elsewhere. For instance, in a medical context 
it is known that certain combinations of drugs when given together cause adverse reactions and should hence not be given  to a patient at the same time.


We now want to obtain the composition of these three diagrams in such a way that the known underlying conflicts are taken into account. 
To the best of our knowledge the only automated approach that can detect the conflicts in the scenarios above given such additional constraints is our work in 
\cite{BC2016}. We now extend our approach to find valid paths in a composed model that avoids these conflicts.

Clearly, to avoid the conflicts the easiest thing to do is to take the second alternative in {\tt sdA} assuming that {\tt pro2} holds. No conflict is present in that case. However, it may be the case that {\tt pro1} holds as well and it has an associated higher priority (preference) leading to the execution of {\tt ma1} followed by {\tt ma2}.   
The question is whether we can still obtain a valid trace that includes this preference and avoids the known message conflicts. Our approach developed here gives an answer to this question under the assumption that simultaneous occurrence of conflicting messages is avoided. Notions of current state, pace and occurrence priority are used 
as parameters to find valid traces in a composed model. We describe how these are treated formally in the next sections. In this paper, we focus on the formal semantics, the composition and valid traces defined at that level, and the formal methods used to detect them. We do not come back to a design level, but we assume the underlying formal models used here have been generated from scenarios or process descriptions. See our earlier work for an idea of the transformation defined at the metamodel level \cite{BABC2015,BBA2015,BBA2016}. See \cite{KB2016} for a description of the medical problem of treating patients with multimorbidities.

\section{Formal Model}
\label{LES-sec}

The model we use to capture the semantics of a sequence diagram is a
labelled (prime) event structure \cite{WinNie95}, or event structure for short. 
The advantages of an event structure are the underlying simplicity of the model and how it naturally describes fundamental notions present in behavioural models including sequential, parallel and iterative behaviour (or the unfoldings thereof) as well as nondeterminism (cf. \cite{Kue-TCS-06,Bow06}). 

In an event structure, we have a set of event occurrences together with binary relations for expressing causal dependency (called \emph{causality}) and nondeterminism (called \emph{conflict}). 
The causality relation implies a (partial) order among event occurrences, while the conflict relation expresses how the occurrence of certain events excludes the occurrence of others (e.g., an event occurring in one operand of an alternative fragment excludes events in another operand).
From the two relations defined on the set of events, a further relation is derived, namely the \emph{concurrency} relation $co$. 
Two events are concurrent if and only if they are completely unrelated, i.e., neither related by causality nor by conflict.
As a derived notion we thus obtain a way to model events associated to locations from different operands in a parallel fragment.
The formal definition, as provided for instance in \cite{Kue-TCS-06}, is as follows.

\begin{defn}
\label{RefDefEs}
\label{ES-def}
An \emph{event structure} is a triple $E = (Ev, \rightarrow^*,\#)$
where $Ev$ is a set of events and $\rightarrow^*, \# \subseteq Ev
\times Ev$ are binary relations called \emph{causality} and \emph{conflict},
respectively. Causality $ \rightarrow^*$ is
a partial order.
Conflict $\#$ is symmetric and irreflexive, and propagates over causality,
i.e.,
$e \# e^{'} \wedge e' \rightarrow^{\ast} e^{''} \Rightarrow e \# e^{''} $ for all
$e, e^{'},e^{''} \in Ev$. Two events $e, e^{'} \in Ev$ are \emph{concurrent},
$e\ co\ e^{'}$ iff $\neg (e \rightarrow^{\ast} e^{'} \vee e^{'}
\rightarrow^{\ast} e \vee e \# e^{'})$.
$C \subseteq Ev$ is a 
\emph{configuration} iff (1) $C$ is \emph{conflict-free}: $\forall e, e' \in C \neg(\ e \# e')$ and (2) \emph{downward-closed}: $e \in C \text{ and } e' \rightarrow^{\ast} e$ implies  $e' \in C$.
\end{defn}

We assume \emph{discrete} event structures.
Discreteness imposes a finiteness constraint on the model, i.e., there are always only a finite number of causally related predecessors to an event, known as the
\emph{local configuration} of the event (written $\dc e$). 
A further motivation for this constraint is  given by the fact that every execution has a starting point or configuration. 
A \emph{trace of execution} in an event structure is a maximal configuration.
An event $e$ may have an immediate successor $e'$ according to the order $\to^*$: 
in this case, we will usually write $e \to e'$.
The relation given by $\to$ is called \emph{immediate causality}.

Event structures are
enriched with a labelling function 
$\mu:Ev\rightarrow 2^L$ that maps each event onto a subset of elements of
$L$. This labelling function is necessary to establish a connection
between the semantic
model (event structure) and the syntactic model it is describing. The set $L$ of labels in our case 
either denote formulas (constraints over integer variables, e.g., $x > 9$ or $y = 5$), logical propositions (e.g., {\tt pro1})
or actions (e.g., $ma1$). 
 If for an event $e\in Ev$, $\mu(e)$ contains an action $\alpha$, then $e$ denotes the occurrence of that action $\alpha$. If $\mu(e)$ contains a formula or logical proposition $\varphi$ then $\varphi$ must hold when $e$ occurs. 

We consider an additional labelling function $\nu:Ev\rightarrow \Nat\times \Nat$  to associate to each event its \emph{priority}
and \emph{duration}. For an event $e$ with $\nu(e)=(p,\_)$, the highest the value of $p$ the higher the priority associated to the event.
The second component of $\nu(e)=(\_,d)$ gives $d$, the time units spent at event $e$. 
The labelling function $\nu$ is used later when fine-tuning the composition with respect to label conflicts. 

%

A labelled event structure over a set of labels $L$ is a triple $M = (Ev,\mu,\nu)$. 
Let $M_1,\dots, M_n$ with $M_i=(E_i,\mu_i,\nu_i)$ 
a finite set of labelled event structures over sets of labels $L_i$ with $1\leq i\leq n$ . Let $\mathcal L=\bigcup_{i=1}^{n} L_i$.
A finite set of \emph{label constraints} defined over $\mathcal L$ is given by $\Gamma\subseteq L_i\times L_j$ where $i\not=j$ characterising label conflicts. 

We do not show how to generate an event structure from a sequence diagram, just the general idea.
The locations along the lifelines of sequence diagrams are associated to one or more events. Locations within different operands of an alternative fragment correspond to events in conflict, whereas locations within operands of  a parallel fragment correspond to concurrent events. The events associated to the locations along a lifeline are related by causality (partial order). For more details see for instance \cite{Kue-TCS-06}.

\begin{figure}[h]
\centering{}
\includegraphics[scale=.33]{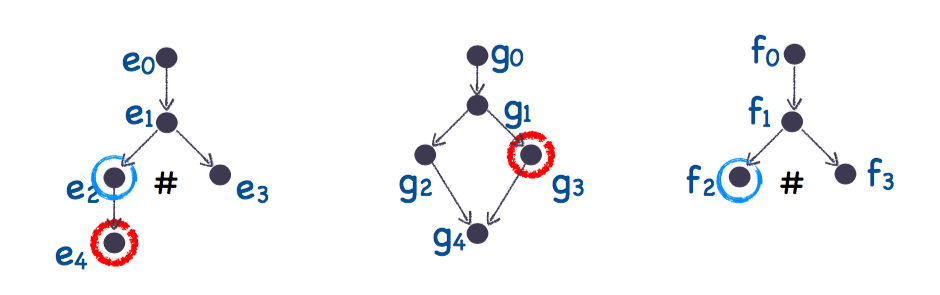}
\caption{Corresponding event structures for instance {\tt p}.}
\label{les-fig}
\end{figure}

Recall the example of Fig.~\ref{sdia-ex1-fig} introduced in the previous section. 
The locations along the lifeline of instance {\tt p} have been marked in Fig.~\ref{sdia-ex1-fig}.
The locations associated to the conditions/guards of the alternative fragments belong to the instances of class {\tt D}, but that distinction is irrelevant for our purposes. 
The label conflicts are given by $\Gamma=\{(ma1,mc1),(ma2,mb2)\}$.
The behaviour of {\tt p} in the individual diagrams of Fig.~\ref{sdia-ex1-fig} is shown in the three event structures $M_A$, $M_B$ and $M_C$ of
Fig.~\ref{les-fig}, where the events are associated to the marked locations of the corresponding sequence diagram as expected. 
The defined labels are as follows: $\mu_A(e_2)=\{pro1,ma1\}$, $\mu_A(e_3)=\{pro2,ma3\}$, and $\mu_A(e_4)=\{ma2\}$ for the event structure associated to {\bf \tt sdA};
$\mu_B(g_2)=\{mb1\}$ and $\mu_B(g_3)=\{mb2\}$ associated to {\bf\tt sdB};
and $\mu_C(f_2)=\{x>9,mc1\}$ and 
$\mu_C(f_3)=\{x>20,mc3\}$ associated to {\bf \tt sdC}.
  


The labels of some of the events above are inconsistent/conflicting according to $\Gamma$, namely
events $e_2$ and $f_2$, and events $e_4$ and $g_3$.
When obtaining the composition of the models above we need to make sure the label inconsistencies are detected and avoided. A composed model that avoids the labels 
could reduce the composition to the trace of execution $\tau_1=\{e_0,e_1,e_3,g_0,g_1,g_2,g_3,g_4,f_0,f_1,f_3\}$ or $\tau_2$ (identical to $\tau_1$ except that it contains $f_2$ instead of $f_3$). However, these traces may not be the best traces of execution. The labels on events are only inconsistent if they occur simultaneously, and if we know where instance {\tt p} is within each of the scenarios we may be able to avoid it. The labelling function $\nu$ gives us that information. 

\begin{figure}[h]
\centering{}
\includegraphics[scale=.375]{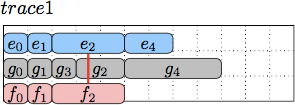}
\includegraphics[scale=.375]{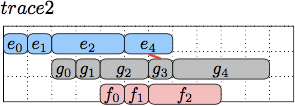}
\includegraphics[scale=.375]{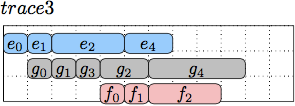}
\caption{Possible traces of execution with and without inconsistencies.}
\label{traces-fig}
\end{figure}

Assume the following $\nu$ labels for some of the events in our example: $\nu(e_0)=\nu(g_0)=\nu(f_0)=(1,1)$, $\nu(e_1)=\nu(g_1)=\nu(f_1)=(1,1)$,
$\nu(e_2)=(5,3)$, $\nu(e_3)=(1,3)$, $\nu(e_4)=(5,2)$, $\nu(g_2)=(1,2)$, $\nu(g_3)=(1,1)$, $\nu(f_1)=(1,1)$, $\nu(f_2)=(3,3)$ and $\nu(f_3)=(1,2)$. Consider 
the possible  traces of execution shown in Fig.~\ref{traces-fig} with time evolving from the left to the right, and considering the events in {\tt sdA} with highest priority (here assumed to have value 5).


The traces illustrate how the event duration and the (dephased) order in which execution is done for the different scenarios may or may not contain inconsistencies. The first two example traces contain inconsistencies, because events with label conflicts occur at the same time. A resolution for $trace 1$ 
could replace the occurrence of $f_2$ with $f_3$ (compromising on the effectiveness of $f_2$ but guaranteeing the higher priority of $e_2$), and for $trace 2$ 
could change the order of occurrence of $g_2$ and $g_3$. Note that when having a conflict between two events with an assigned priority we always try to satisfy the event with the highest priority first. Here $e_2$ has priority $5$ and $f_2$ has priority $3$, so we favour $e_2$. If both events had the same priority the resolution would pick one of the events at random.
In  $trace 3$ no inconsistencies are present and all events have the highest priority.
In the next section we show how we can generate automatically the valid traces for a set of labelled event structures given a set of label conflicts and 
the degree that each structure is being dephased.
%

%

\section{\I{} and Z3 Combined}
\label{iz3-sec}

\newcommand{\modulator}{f}
\newcommand{\interactionDatabase}{D}
\newcommand{\scoreDatabase}{d}
\newcommand{\N}{\mathbb{N}}

We combine two formal techniques to \emph{calculate automatically the outcome} of the composition of two or more behavioural models as a set of allowed traces and to determine that the \emph{result is correct}: the theorem prover \I{} \cite{Isabelle-HOL} and the SMT solver Z3 \cite{z3}.

\I{}  is a theorem prover or proof assistant which provides a framework to accommodate logical systems (deductive rules, axioms), and compute the validity of logical deductions according to a given system.
In this paper, we use Isabelle's library based on \emph{\hol{}} (HOL).
In HOL, the basic notions are type specification, function application, lambda abstraction, and equality.
In addition to be able to check logical inference over logical systems, theorem provers such as Isabelle also contain
automated deduction tools, and interfaces to external tools such as SMT solvers and automated theorem provers. We use the theorem prover to guarantee the correctness of our models, the composition result and traces.

A satisfiability modulo theories (SMT) solver is a computer program designed to check the satisfiability of a set of formulas (known as \emph{assertions}) expressed in first-order logic, where for instance arithmetic operations and comparison are understood, and additional relations and functions can be given a semantic meaning
in order to make the problem satisfiable.
Within proof assistants, SMT solvers are used to find proofs by adding already proved theorems to the list of assertions, and by negating the statement to be proved to reach a contradiction. If a SMT solver returns {\sf{unsat}}, then a proof can be reconstructed from the given assertions.
The integration between \I{} and SMT solvers such as Z3 provides users an additional powerful combination to
be able to produce more proofs automatically. We use the SMT solver to identify label inconsistencies, which may require the use of arithmetic operations and comparison, and to find a solution which avoids the inconsistencies and considers the additional labelling information given by $\nu$.

Let $ M_1, \ldots, M_n $, with $M_i=((Ev_i, \rightarrow_i^*,\#_i),\mu_i,\nu_i)$ and $1\leq i\leq n$ over a set of labels $L_i$, be a list of finite event structures. Let $\Gamma\subseteq L_i\times L_j$
with $i\not = j$ denote the set of label conflicts.
We assume that the corresponding sets of events 
are pairwise disjoint. In what follows we denote the immediate causality $\rightarrow_i$ by $G_i$, and set
\begin{align*}
G := \bigcup_{i=1,\ldots, n} G_i,
\\
\# := \bigcup_{i=1, \ldots, n} \#_i.
\end{align*}
Given a relation $R$ over a set $Y$ and a set $X \subseteq Y$, we introduce the notation $\map{R}  \left( X \right)$ to denote the image of $X$ through $R$.

We will now proceed in steps: first, we show how to compute traces, then we show how to use $\nu$ to obtain the preferred one, depending on the duration and priority assigned to single events.
In doing so, we will write formulas close to the first-order logic language used by SMT solvers; for the sake of readability, however, we will employ some simplifications.
In particular, we adopt infix notation instead of prefix notation, we use set-theoretical styling instead of predicates (e.g., writing $ \left( j, k \right) \in G_i$ in lieu of $ G_i\  j\  k$), and we omit type specifications.

\subsection{Trace Calculation}
\label{RefSectTrace}
To represent an execution trace, we need to express which events are part of it, and in which order.
The first piece of information will be given by a boolean function over all the events, namely $\sel$.


We can compute $\sel$ using an SMT solver as follows.
Let us illustrate the procedure for a fixed event structure $Ev_i$.
The conditions of $\sel$ being conflict-free and downward-closed (see Definition~\ref{RefDefEs}) 
are straightforward to express:

\begin{align*}
\forall j, k \in Ev_i. \ \sel \left( j \right) \wedge \sel \left( k \right) \rightarrow \neg \left( j \# k  \right)
\end{align*}

\begin{align*}
\forall j \in \ran \left( G_i \right). \ 
\sel \left( j  \right) \rightarrow \bigwedge_{k \in \map{\left( G_i^{-1} \right)} \left\{ j \right\}} \sel \left(  k  \right)
\end{align*}

The two formulas above capture the notion of configuration (see Definition~\ref{RefDefEs}) in a way amenable to SMT solvers.
To compute traces of execution (Section~\ref{LES-sec}), we have to capture the notion of a \emph{maximal} configuration.
This notion implies quantifying over configurations, which is not allowed in the first-order logic universe of SMT solvers: sets in general are not first-class objects.
However, the notion of maximality can be reformulated in the case of configurations of finite event structures as follows:

\begin{align}
\label{RefFormulaMaximalitySMT}
\forall z \in Ev_i. \ 
\exists y \in Ev_i. 
&&
(
\left( y \# z \wedge \sel \left( y  \right)  \right)
\vee
\\
\notag{}
&&
\left(  \left( y, z \right) \in G_i \wedge \neg \sel \left( y \right)  \right)
).
\end{align}


The formulas above can be used to compute traces via an SMT solver; 
more precisely, the events for which $\sel$ is true represent the event set of a trace, and the event set of any legal trace satisfies the assertions above.
We will formally prove the correctness of this statement in Section~\ref{ver-sec}.

To add an order to this set, we proceed as follows.
First, taken a single $G_i$, we need to obtain the corresponding partial order $P_i$ (effectively obtaining the original causality relation $\rightarrow_i^{\ast}$), which can be derived 
from 
the following assertions:

\begin{align*}
\forall j,\  k. \left( j, k \right) \in G_i \to \left( j, k \right) \in P_i,
\\
\forall j, \  k,\ l. \left( j, k \right) \in P_i \wedge \left( k, l \right) \in P_i \to \left( j, l \right) \in P_i,
\\
\forall j \in Ev_i. \left( j, j \right) \in P_i,
\\
\forall j,\ k. \left( j, k \right) \in P_i \wedge \left( k, j \right) \in P_i \to j=k.
\end{align*}

We now use $P_i$ to obtain a sorting of all the selected events of $Ev_i$.
This can be done by introducing an injective function $s_i : Ev_i \to \N$, and then imposing that it is order-preserving (between the partial order $P_i$ and the canonical order relation for natural numbers), surjective over the integer interval $ \left[ 1, \ldots, \left| Ev_i \right| \right]$, and such that $s_i \left( j \right) < s_i \left( k \right)$ whenever $j$ is selected and $k$ is not:

\begin{align*}
\forall j,\ k. \left( j, k \right) \in P_i \to s_i \left( j \right) \leq s_j \left( k \right),
\\
\forall j,\ k \in Ev_i. j \neq k \to s_i \left( j \right) \neq s_i \left( k \right),
\\
\forall j \in Ev_i. s_i \left( j \right) \geq 1
\\
\forall j \in Ev_i. s_i \left( j \right) \leq \left| Ev_i \right|
\\
\forall j,\ k \in Ev_i. \sel \left( j \right) \wedge \neg \sel \left( k \right) \to s_i \left( j \right) < s_i \left( k \right).
\end{align*}

\subsection{Using $\nu$ for Trace Selection}
\label{RefSectTraceSelect}
As done in the example of Fig.~\ref{traces-fig}, we want to be able to determine whether events from distinct event structures overlap, in order to decide whether the conflict they might have is triggered or not.
We associate a $\clock$ function to each event, expressing the time when the event starts.
To calculate it, we use the sorting functions $s_i$ obtained in the previous section, together with the duration of each event provided by $\nu$.
This can be done by requiring that an event following another (according to $s_i$) starts exactly when the latter ends:
\begin{align*}
\forall j,\ k \in Ev_i. 
\\
\left( \sel j \wedge \sel k 
\wedge s_i \left( j \right) \leq \left| Ev_i \right| \wedge s_i \left( k \right) \leq \left| Ev_i \right|
\wedge s_i \left( k \right) - s_i \left( j \right) = 1
\right)
\\
\to 
\clock \left( k \right) = \clock \left( j \right) + \nu_2 \left( j \right), 
\end{align*}
where $\nu_2$ is the second component of $\nu$, yielding the duration.

The formula above leaves the clocks of the roots undetermined, hence we need to set them separately.
This allows us to introduce dephasing between different models, by specifying different clocks for the roots of different models, which means starting each model at dephased times.
Finally, the concept of clock allows us to avoid inconsistencies due to events mutually in conflict, but whose occurrence is not simultaneous.

To attain this goal, we assign a priority (which we also refer to as score) to each event and to each pair of events from distinct models, through the function $\score$ and $\Score$, respectively, both yielding integer values.
$\Score \left( j, k \right)$ will take into account both the absolute conflict between events $j$ and $k$, and their clock, in order to decide whether they are in conflict given a trace (recall, from the definition above and the definition of $s_i$ in previous section, that each trace determines clock values for each event).
Formally, this is obtained by the following requirement, repeated for all $ m \neq n$, $ m, n \in \left\{ 1, \ldots, n \right\}$:
\begin{align*}
\forall j \in Ev_m ,\  k \in Ev_n. \sel \left( j  \right) \wedge \sel \left( k  \right) 
\rightarrow{} 
\Score \left( j, k \right) =
\\
\modulator \left(
\clock \left( j \right), 
\clock \left( k \right), 
\nu_2 \left( j \right),
\interactionDatabase \left( \mu \left( j \right), \mu \left( k \right)  \right) 
 \right),
\end{align*}
where $\interactionDatabase$ calculates the absolute conflict (a negative number) between events based on their label, and is passed to $\modulator$. Further, $\modulator$ combines that with the distance of the event occurrences to obtain the effective result, as follows:

\begin{align*}
\modulator \left( x_1, x_2, y, z \right) :=
\left\{  
\begin{aligned}
& z, \text{ if } x_2 - x_1 \in [0,y]
\\
& 0, \text{ otherwise. }
\end{aligned}
\right.
\end{align*}

Besides conflicts between events in distinct models, the other criterion when picking a trace is the absolute priority of each event.
Therefore, we also require
\begin{align*}
\forall j. \sel \left( j \right) \to \score \left( j \right) = \nu_1 \left( j \right)
\\
\forall j. \neg \sel \left( j \right) \to \score \left( j \right) = 0,
\end{align*}
where $\nu_1$ is the first component of $\nu$, yielding the priority.

To obtain the final trace, we sum over all the $\Score \left( j, k \right)$ and over all the values $\score \left( j \right)$, and pick the trace maximising such sum.
To do so, we need to exploit the optimizing part of the SMT solver Z3, $\nu$Z~\cite{bjorner2015nuz}.

\subsection{Example}

We test the output of our approach with respect to the simple example of Fig.~\ref{traces-fig}.
In the first case ($trace 1$ of Fig.~\ref{traces-fig}), all the models start executing together, 
and the SMT solver yields the optimal trace on the left of Table~\ref{RefTable}.
The incompatibility between $g_3$ and $e_4$ does not pose problems, since those two events cannot overlap.
However, the solver has been forced to choose between the branch starting at $e_2$ and $f_2$.
Given that the $e_2$ branch has the highest priority overall, it has been picked.
But event $f_2$ also has a high priority, which leads to his choice over $f_3$, as soon as dephasing allows that.
We now test that this is indeed the case.
The right-hand side of Table~\ref{RefTable} displays
the output resulting from running the same experiment, but with $f_0$ happening at time $4$ and $g_0$ happening at time $1$ (corresponding to $trace 3$ in Fig.~\ref{traces-fig}):

\begin{table}
\caption{Outputs corresponding to $trace 1$ (left) and $trace 3$ (right)}
\label{RefTable}
\begin{small}
\begin{tabular}[h]{|c|c|c|c|c|}
\hline
clock & event & order & priority & duration
\\
\hline
0 & e0 & 1 & 1 & 1 
\\
\hline
0 & f0 & 1 & 1 & 1 
\\
\hline
0 & g0 & 1 & 1 & 1 
\\
\hline
1 & e1 & 2 & 1 & 1 
\\
\hline
1 & f1 & 2 & 1 & 1 
\\
\hline
1 & g1 & 2 & 1 & 1 
\\
\hline
2 & e2 & 3 & 5 & 3 
\\
\hline
2 & f3 & 3 & 1 & 2 
\\
\hline
2 & g2 & 3 & 1 & 2 
\\
\hline
4 & g3 & 4 & 1 & 1 
\\
\hline
5 & e4 & 4 & 5 & 2 
\\
\hline
5 & g4 & 5 & 1 & 4 
\\
\hline
\end{tabular}
\hfill{}
\begin{tabular}{|c|c|c|c|c|}
\hline
clock & event & order & priority & duration
\\
\hline
0 & e0 & 1 & 1 & 1 
\\
\hline
1 & e1 & 2 & 1 & 1 
\\
\hline
1 & g0 & 1 & 1 & 1 
\\
\hline
2 & e2 & 3 & 5 & 3 
\\
\hline
2 & g1 & 2 & 1 & 1 
\\
\hline
3 & g3 & 3 & 1 & 1 
\\
\hline
4 & f0 & 1 & 1 & 1 
\\
\hline
4 & g2 & 4 & 1 & 2 
\\
\hline
5 & e4 & 4 & 5 & 2 
\\
\hline
5 & f1 & 2 & 1 & 1 
\\
\hline
6 & f2 & 3 & 3 & 3 
\\
\hline
6 & g4 & 5 & 1 & 4 
\\
\hline
\end{tabular}
\end{small}
\end{table}

Now, the incompatibility between $e_2$ and $f_2$ can be avoided by dephasing, and indeed both events are part of the new trace.
We also note that the incompatibility between $e_4$ and $g_3$ has also been avoided by swapping the execution of $g_2$ and $g_3$, as expected.

\section{Verification}
\label{ver-sec}

The first-order language used in SMT solvers often requires laborious and error-prone translation from higher-level mathematical abstractions.
Let us take the notion of event structure as an example: the concepts of partial order, and relation in general are expressed typically through sets of ordered pairs.
However, the notion of set is not directly available in SMT-LIB, and one is forced to choose a lower-level representation of it.
For example, by representing relations as boolean predicates taking two arguments; this, in turn, typically makes higher-level operations (such as composition, image, taking the domain, injectivity, etc) more complicated.

A way of tackling the complexity arising from this translation, and to make sure that it correctly represents the involved objects, is to write the wanted original definitions in a higher-order  language (for example higher-order logic, HOL) which allows to express them easily.
In the same language, we can of course also write definitions closer to the ones required for SMT solvers.
The crucial point is that \I{} provides an SMT-LIB generator which can generate, from the latter definitions, SMT assertion directly executable by SMT solvers.
And, at the same time, we can prove, inside \I{}, the equivalence between the standard definitions and those closer to the SMT language. 
Since the latter directly generate the SMT code used for our computations, the formal equivalence proof is also a proof of correctness for our generated SMT code.

Hence, we write in \I{} a definition of event structure which is close to Definition~\ref{RefDefEs}:
\begin{lstlisting}
abbreviation "isLes causality conflict == 
propagation conflict causality & sym conflict & 
irrefl conflict & trans causality & 
antisym causality & reflex causality".
\end{lstlisting}

Above 
{\tt isLes causality conflict} returns \lstinline!true! exactly when \lstinline!causality! and \lstinline!conflict! constitute a valid event structure.
In the definition above, causality and conflicts are sets of pairs, which permits to use the standard property of symmetry (\lstinline!sym!), transitivity (\lstinline!trans!) already present in the \I{} libraries.
We only needed to introduce \lstinline!propagation! as a direct translation of the propagation condition occurring in Definition~\ref{RefDefEs}, which we omit here.

On the other hand, an equivalent definition is also introduced in \I{}:
\begin{lstlisting}
abbreviation "IsLes Causality Conflict == 
Propagation Conflict Causality & Sym Conflict & 
Irrefl Conflict & Trans Causality & 
Antisym Causality & Reflex Causality",
\end{lstlisting}
which is similar to the previous one, but where 
{\tt Causality} and 
{\tt Conflict} are no longer sets, but predicates.

This allows us to use the definition of 
{\tt IsLes}
 for producing SMT code directly through \I{}'s SMT generator.
Since this generator is originally provided for theorem proving, and not for direct SMT computations as we are interested here, we have to trick \I{} into proving a lemma:

\begin{lstlisting}
lemma assumes "IsLes Causality Conflict" shows False 
sledgehammer run [provers=z3, minimize=false, 
		  overlord=true, timeout=1] (assms)
\end{lstlisting}

The lemma above makes some assumptions (hypotheses) written after the keyword \texttt{assumes}. 
The assumptions include that the two relations described constitute a valid event structure.
The keyword \verb|shows| introduces the thesis (here {\tt False}) and
\verb|sledgehammer| is \I{}'s command for referencing outside tools (ATPs, SMT solvers), used here to run Z3. 
We note that the argument \texttt{assms} is used to instruct \texttt{sledgehammer} to ignore any other theorems in the Isabelle library and consider only the stated assumptions.

In the  lines above, Isabelle  will pass to \z{} a file which contains one declaration for each of the relations \verb|Causality| and \verb|Conflict|, and assertions for each of the stated hypotheses.
In the present case, we only have one hypothesis, which will result in an SMT definition of event structure, directly usable for our computations.

The last step to certify the correctness of this SMT generated code is to prove the equivalence of 
{\tt isLes} and 
{\tt IsLes}, which is attained through the following theorem:

\begin{lstlisting}
theorem "IsLes causality conflict $\leftrightarrow$ 
	(isLes (pred2set Causality) (pred2set Conflict))",
\end{lstlisting}
where \lstinline!pred2set! converts from relations represented as predicates into relations represented as sets.

\begin{figure}[t]
\centering
\includegraphics[width=.6\linewidth]{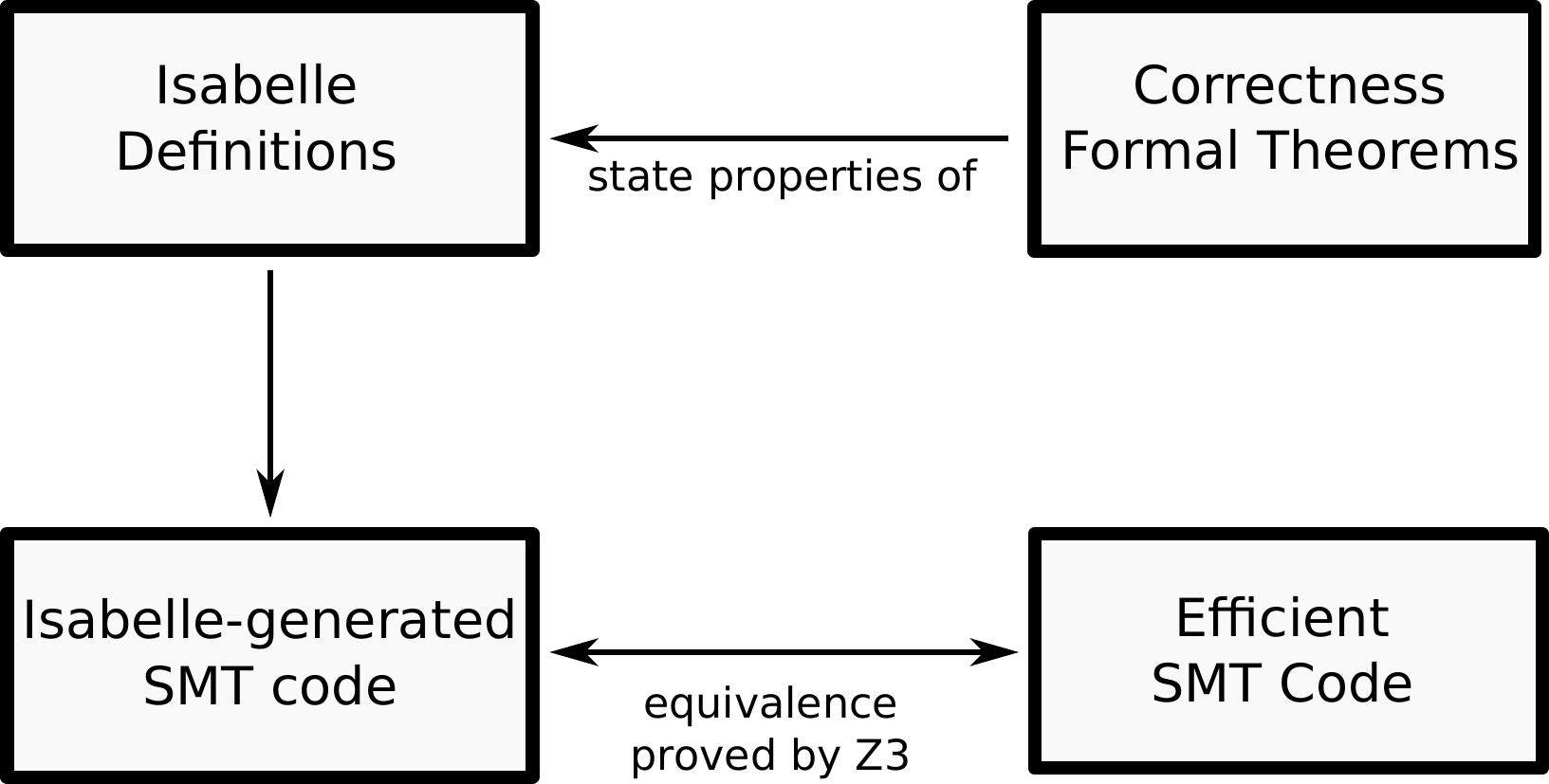}
\caption{Overview of the formal verification of the SMT code.}
\label{RefFigDiagram}
\end{figure}

The idea of using \I{} as an interface to SMT code becomes even more fruitful in cases where the SMT code used for computing a given object departs substantially from the original or standard mathematical definition of that object.
This usually happens, e.g., because the original definition is not directly expressible as a finite number of formula in first-order logic (the language of SMT solvers), or because, even if it is, it is inefficient.
In such cases, we can express both the original definition and the definition used for SMT computing in \I{}, which we can then use both to generate the SMT code for the latter and to formally prove the equivalence of the two definitions, as from the diagram in Figure~\ref{RefFigDiagram}.
As an example, let us take the trace computation seen at the beginning of Section~\ref{RefSectTrace}: there we had to resort to an alternative, less intelligible definition of maximality of configuration \eqref{RefFormulaMaximalitySMT}, because the original definition implied quantifying over all configurations.

In \I{}, we can easily render the pen-and-paper definitions of event structure (which we have seen earlier), of configuration and of trace.
We start by writing the condition specifying that our candidate configuration \lstinline!C! is conflict-free:

\begin{lstlisting}
definition "isConflictFree Cf C = ((C $\times$ C) $\cap$ Cf = {})",
\end{lstlisting}
and the condition about \lstinline!C! being downward closed:

\begin{lstlisting}
definition "isDownwardClosed Ca C = 
	(C $\subseteq$ events Ca & 
	($\forall$ e f. e $\in$ C & (f, e) $\in$ Ca $\rightarrow$ f $\in$ C))".
\end{lstlisting}

This allows the immediate definition of configuration:

\begin{lstlisting}
definition "isConfiguration Ca Cf C = 
  isConflictFree Cf C & isDownwardClosed Ca C",
\end{lstlisting}
\noindent{}
and that of being a trace:

\begin{lstlisting}
definition "isTrace Ca Cf C = 
  isConfiguration Ca Cf C & 
  ($\forall$ Y. Y $\supset$ C $\rightarrow$ $\neg$ (isConfiguration Ca Cf Y))",
\end{lstlisting}
where the last line expresses the maximality of the configuration \lstinline!C!.
We write the same line (i.e., maximality) in the way seen in Section~\ref{RefSectTrace}:

\begin{lstlisting}
abbreviation "isMaximalConfSmt Ca Cf C == 
  ($\forall$ z $\in$ events Ca - C. 
     z $\in$ Cf``C $\vee$ (immediatePredecessors Ca {z})-C $\neq$ {})",
\end{lstlisting}
where \lstinline!immediatePredecessors Ca {z}! returns all the events $e$ satisfying $e \to z$ (we recall that $\to$ is the immediate causality obtained from $\to^*$).
Finally, the following \I{} theorem states that \eqref{RefFormulaMaximalitySMT} is equivalent, for a configuration of a finite event structure, to the original trace definition: 

\begin{lstlisting}
theorem correctness: assumes "finite Ca" "isLes Ca Cf" 
 "isConfiguration Ca Cf C" shows 
 "(isTrace Ca Cf C) $\leftrightarrow$ isMaximalConfSmt Ca Cf C"
\end{lstlisting}

We note that the theorem assumes that \lstinline!C! is a configuration: 
this is not a problem because, as seen in Section~\ref{RefSectTrace}, the notion of configuration admits a straightforward formulation in SMT, while the problematic one is that of \emph{maximality} for a configuration.
We also note that \lstinline!isMaximalConfSmt! builds on \lstinline!immediatePredecessors Ca!, rather than directly on \lstinline!Ca!.
This is also not a problem, since the SMT computations we introduced in Section~\ref{RefSectTrace} take as input the immediate causality relations $G_i, i=1, \ldots, n$, and use them to calculate via SMT the causalities $\to_i^*$.

The \I{} definition \lstinline!isMaximalConfSmt! can be used to automatically generate SMT code through \lstinline!sledgehammer!, as we did with \lstinline!IsLes!.
This corresponds to the vertical arrow on the left in Diagram~\ref{RefFigDiagram}.
In this case, however, the obtained SMT code is not as efficient as the one we manually wrote in Section~\ref{RefSectTrace}: it is a general fact that the efficiency of SMT code can depend dramatically on formal details, such as eliminating quantifiers by explicit enumeration, rewriting the assertions in normal forms, etc\ldots{}
We want to keep both the efficiency of the manually-written SMT code and the correctness of the \I{}-generated SMT code.
Our solution is to take both, and prove their equivalence using the SMT solver itself.
This corresponds to the horizontal arrow at the bottom of Diagram~\ref{RefFigDiagram}, and can be implemented as follows.
We introduce an SMT boolean function \lstinline!maximality! which is true exactly when \eqref{RefFormulaMaximalitySMT}, repeated for each $i=1,\ldots,n$, is true.
We also introduce another boolean function \lstinline!maximalityIsabelle!, defined by using the SMT code generated by \I{} using \lstinline!isMaximalConfSmt!.
If \lstinline!maximality! and \lstinline!maximalityIsabelle! were not equivalent, there would be some $\sel$ satisfying one but not the other.
Therefore, we challenged the SMT solver as follows:

\begin{lstlisting}
(assert (or (and maximality (not maximalityIsabelle)) 
	    (and (not maximality) maximalityIsabelle))),
\end{lstlisting}
obtaining the answer \lstinline!(unsat)!, 
which guarantees that the SMT code we use for trace maximality calculation is correct.
Correctness, as usually, means that if we trust the SMT solver, \I{}, and the environment in which they run, then we can trust that the result of our computation is indeed a trace.
Not only: we can rest assured that any trace will satisfy the SMT formula (i.e., Formula~\ref{RefFormulaMaximalitySMT}) passed to the solver for the computation.
To increase our confidence in the results, we could also prove the correctness of the remaining computations, i.e., the trace selection (Section~\ref{RefSectTrace}).
The general mechanism represented in Figure~\ref{RefFigDiagram} could again be applied: we would need to write an \I{} formal specification of the desired property guiding the trace selection, write an \I{} definition to generate SMT code, and an \I{} theorem proving that the latter obeys the former.
Finally, we would use the SMT solver to prove that the \I{}-generated SMT code and the manually written SMT code are equivalent.
Again, this would imply correctness as soon as we trust the solver and \I{}; additionally, in this case we would also need to trust $\nu$Z (see end of Section~\ref{RefSectTraceSelect}), which is not used in trace computation but only in trace selection.

\section{Related Work}\label{rel-sec}

Systems are usually designed through a combination of several models, some to capture structural aspects and some to describe more complex aspects of behaviour. 
As argued in \cite{HarMar03-book}, modelling the complete behaviour of a component or subsystem is difficult and error prone. Instead, it is easier to formulate partial behaviour as scenarios in Live Sequence Charts (LSCs), UML sequence diagrams or similar. One of the problems that arises from partial modelling is potentially inconsistent or incomplete behaviour. 

When looking at the integration of several model views or diagrams, Widl et al. \cite{widl2012guided} deal with composing concurrently evolved sequence diagrams in accordance to the overall behaviour given in state machines. They make direct use of SAT-solvers for the composition. Liang et al.~\cite{liang2008general} present a method of integrating sequence diagrams based on the formalisation of sequence diagrams as typed graphs. Both these papers focus on less complex structures. For example, they do not deal with combined fragments, which can potentially cause substantial complexity.
Bowles and Bordbar \cite{bowles2007formalAndBordbar} present a method of mapping a design consisting of class diagrams, OCL constraints and sequence diagrams into a mathematical model for detecting and analysing inconsistencies. It uses the same underlying categorical construction as done in \cite{Bow06} but it has not been automated.
On the other hand, Zhang et al. \cite{zhang2009} and Rubin et al. \cite{rubin2008} use Alloy for the composition of class diagrams. They transform UML class diagrams into Alloy and compose them automatically. 
They focus on composing static models and the composition code is produced manually. 

We used Alloy to automatically compose sequence diagrams in \cite{BABC2015,BBA2015}. 
Our experience with Alloy has shown that  
it has limitations which have a direct  impact on the scalability of the approach \cite{BBA2016}.
There is an exponential growth in time when trying to compose diagrams with an increasing number of elements, which becomes unusable in practice. The Alloy analyzer is SAT solver-based and SAT- solving time may increase enormously, depending on factors such as the number of variables and the average length of the clause \cite{HotCore}. Z3 \cite{z3} performs much better and we have used it in more recent work \cite{BBA2016,KB2016,BC2016}. We do not know of other approaches using Z3 for model composition. 

We are addressing inconsistent combination of behavioural models in this paper. 
A SAT-based approach, such as Alloy, would allow us to detect inconsistencies and highlight them, as a result of not being able to generate a solution for the composition.
When two or more scenarios combined have inconsistencies, a designer benefits not only from knowing which inconsistencies there are, but what traces of execution can bypass the inconsistencies. 
In practice, it is unlikely that inconsistencies can be removed altogether, and instead we want to find the traces that are valid, avoid the inconsistencies, and may satisfy additional criteria such as priorities. 
SAT solvers cannot be used in this case whereas we have shown that SMT solvers can  in another context \cite{KB2016}. 
The present paper makes a novel contribution by showing how SMT solvers such as Z3 can be used to find the best solution to a generally unsolvable problem of composing models with known inconsistencies.
Finally, the typical combination of SMT solvers and proof assistants is done to help finding proofs, and we bring this combination into a completely different setting for detecting and resolving problems in complex behaviour. 

\section{Conclusions}\label{con-sec}

Inspired by a problem from the medical domain, we have explored a novel approach to compose scenarios and their underlying, possibly dephased, traces of execution. 
Our approach allows us to detect and avoid inconsistencies (if possible) to generate a valid set of traces of execution for a composed model. 
The traces can be fine-tuned to take into account additional requirements on the degree of priority that one model or certain steps in a process (events in our approach) have over other models or alternatives. 
Moreover, our approach is able to find the best trace of execution with respect to these constraints. 
Key to our approach is the use of SMT solvers to search for the best solution. 
Our approach uses a novel combination of the theorem prover Isabelle and the constraint solver Z3, where the theorem prover is fundamental to guarantee the correctness of the approach and to facilitate the interaction with Z3 through the provided SMT-LIB generator. 
This is important because, on one hand, writing SMT code directly is time-consuming and error-prone while, on the other hand, the existing interfaces of SMT solvers with higher-level languages (e.g., APIs) are not currently, to the best of our knowledge, formally verified.

This paper focused on the semantics of the underlying behavioural models. 
Separately we are developing mechanisms to visualise the solutions obtained back to the designer. 
We have used Graphviz in our earlier work in \cite{BABC2015,BBA2015} to show the composition solution obtained with Alloy. 
In future work we want to explore visualisations that work directly on the modelling approaches used by designers, and in particular in the case of inconsistencies, can show them more effectively; 
thus we also aim at achieving an increased adoption of our approach by designers, which in turn is needed to test and validate our techniques on realistic application problems.
Work is in progress to generalize the time representation to allow the duration of an event to be a range, rather than a specific amount of time units.
A further direction for future work is to make the scheme presented here to deal with incompatibilities and priorities even more flexible by using soft constraints: currently, the trace selection is performed by expressing a maximisation problem with hard constraints only; however, soft constraints can be implemented, e.g., via the SMT-LIB command \lstinline|check-sat-assuming|.
Finally, future work will also tackle the issue of finding a way to accommodate indefinite loops and non-terminating behaviours, possibly present in given models, in our approach.

\end{document}